\documentclass[a4paper,11pt]{article}

\usepackage[utf8x]{inputenc}
\usepackage[T1]{fontenc}
\usepackage{color}
\usepackage{amssymb}
\usepackage{amsmath}
\usepackage{graphicx}
\usepackage{mathtools}
\usepackage{ragged2e}
\usepackage{physics}
\usepackage{todonotes}
\usepackage{dcolumn}
\usepackage{bm}
\usepackage{graphicx}
\usepackage{braket}
\usepackage{verbatim} 
\usepackage[english]{babel}
\usepackage{hyperref}
\hypersetup{
citecolor=red,
colorlinks=true,
filecolor=red,
linkcolor=blue,
linktocpage=true,
urlcolor=blue
} 
\usepackage[titletoc,toc,title]{appendix}
\numberwithin{equation}{section}
\usepackage{cleveref}
\usepackage[margin=2.5cm]{geometry}
\usepackage{cite}
\usepackage{epsfig}
\usepackage{float}
\usepackage{amsfonts}
\usepackage{enumitem}
\usepackage[font={footnotesize,it}]{caption}

\numberwithin{equation}{section}

\begin{document}

\begin{center}
\centerline{\Large {\bf Charged Black Hole with String Cloud Deformation: Entanglement and Chaos}}

\vspace{8mm}

\renewcommand\thefootnote{\mbox{$\fnsymbol{footnote}$}}
Shagun Kaushal,${}^{1,2}$\footnote{shagun123@iitd.ac.in, shagun.kaushal@vit.ac.in}
Arpit Maurya,${}^{3}$\footnote{arpit.20phz0009@iitrpr.ac.in}
Sanjay Pant,${}^{4}$\footnote{sanjaypant.phy@geu.ac.in}
Himanshu Parihar ${}^{5, 6}$\footnote{himansp@phys.ncts.ntu.edu.tw}

\vspace{4mm}
${}^1${\small \sl Department of Physics}\\
{\small \sl IIT Delhi}\\
{\small \sl  New Delhi 110016, India} 
\vskip 0.2cm
${}^2${\small \sl Department of Physics}\\
{\small \sl Vellore Institute of Technology}\\
{\small \sl  Vellore 632014, India \footnote{Current Affiliation}}

\vskip 0.2cm

${}^3${\small \sl Department of Physics}\\
{\small \sl IIT Ropar}\\
{\small \sl  Runagar Punjab, India} 

\vskip 0.2cm

${}^4${\small \sl Department of Allied Sciences (Physics)}\\
{\small \sl Graphic Era (Deemed to be University)}\\
{\small \sl  Dehradun, Uttarakhand 248002, India} 

\vskip 0.2cm
${}^5${\small \sl Center of Theory and Computation}\\
{\small \sl National Tsing-Hua University}\\
 {\small \sl Hsinchu 30013, Taiwan} 
 \vskip 0.2cm
${}^6${\small \sl Physics Division}\\
   {\small \sl  National Center for Theoretical Sciences}\\
     {\small \sl Taipei 10617, Taiwan} 
     \vskip 0.2cm
\end{center}

\vspace{6mm}
\numberwithin{equation}{section}
\setcounter{footnote}{0}
\renewcommand\thefootnote{\mbox{\arabic{footnote}}}

\begin{abstract} 
We perform a holographic analysis of several quantum information-theoretic observables—entanglement entropy (EE), mutual information (MI), entanglement wedge cross-section (EWCS), butterfly velocity ($v_B$) and thermo mutual information (TMI)—in the background of charged AdS black hole deformed by a homogeneous string cloud.
This configuration is dual to a large $\mathcal{N}_c$ strongly‑coupled field theory at finite temperature and finite chemical potential, in presence of quark‑cloud. 
 We study how the entanglement structure and chaotic dynamics in the boundary theory are affected by the charge and backreaction. We find that both EE and EWCS increase monotonically with charge and backreaction, indicating enhanced correlations due to additional bulk degrees of freedom. On the other hand MI and TMI show a more intricate dependence: backreaction tends to strengthen correlations, while increasing charge suppresses entanglement and makes the system more susceptible to scrambling. The analysis of the butterfly velocity \(v_B\) indicates that both the presence of charge and the backreaction suppress the chaotic behavior of the system by reducing \(v_B\). Furthermore, TMI exhibits a sharp transition under shockwave perturbations, with inter-boundary entanglement being entirely disrupted beyond a critical shock strength, which decreases with increasing charge. 
\end{abstract}

\newpage
\tableofcontents
\newpage

\section{Introduction}\label{sec:intro}

Strongly coupled quantum field theories are difficult to study because standard perturbative methods which rely on weak interactions, fail to capture essential physical phenomena. These include confinement in QCD, mass gaps in non-Abelian gauge theories and long-range entanglement near quantum critical points, all of which require non-perturbative approaches.  
In strongly coupled regimes, quantum entanglement serves as a key probe of nonlocal correlations, while quantum chaos—characterized by out-of-time-ordered correlators (OTOCs)—provides insight into the spread of information and thermalisation dynamics. Together, entanglement and chaos reveal deep connections between quantum information and dynamics in strongly interacting systems. To study these effects systematically, we need a non-perturbative framework. The AdS/CFT correspondence \cite{Maldacena:1997re} provides such a framework where strongly coupled field theories are mapped to classical gravitational theories in higher dimensions. This duality has opened up new directions in understanding quantum gravity, black hole physics and the entanglement structure of strongly coupled systems.

In quantum information theory, the definition and measurement of entanglement depends on the state in question. For bipartite pure states, the canonical measure is the von Neumann entropy of the reduced density matrix known as the entanglement entropy (EE). This measure is highly effective, not only quantifying entanglement but also revealing crucial aspects like critical behavior and phase transitions in many-body systems. However, for mixed states, entanglement entropy provides an incomplete picture as it captures both quantum entanglement and classical (or thermal) correlations without distinction. To overcome this limitation, various other entanglement measures have been developed to probe the structure of correlations in mixed states. These include mutual information (MI) which provides a symmetric estimate of total correlations, entanglement negativity (EN) \cite{Vidal:2002zz, Plenio:2005cwa} which serves as a partial entanglement monotone and the entanglement of purification (EoP) \cite{Terhal:2002riz} which captures both quantum and classical correlations in a more refined way. Another recent addition is the reflected entropy, offering insight into entanglement in bipartite mixed states \cite{Dutta:2019gen}.

While these measures provide significant conceptual insight, their explicit computation in strongly coupled field theories—especially in higher dimensions—remains highly challenging due to the inherently non-perturbative nature of such systems.
This is where holography provides a powerful computational framework. In the context of AdS/CFT correspondence, many of these entanglement measures admit elegant geometric duals, enabling their study in regimes inaccessible via traditional field theory methods. The AdS/CFT correspondence is a duality between a large $\mathcal{N}_c$, \( d \)-dimensional conformal field theory (CFT\(_d\)) and a gravitational theory in asymptotically AdS\(_{d+1} \) spacetime. This holographic equivalence provides a non-perturbative definition of quantum gravity in AdS in terms of a strongly coupled gauge theory on the boundary \cite{Maldacena:1997re, Witten:1998qj}.
 A key example of this correspondence is the Ryu-Takayanagi (RT) prescription \cite{Ryu:2006bv, Ryu:2006ef} which offers a geometric definition for the entanglement entropy of a spatial region in the dual CFT. According to this proposal, the entanglement entropy is given by 
\begin{equation}
S_A = \frac{\mathrm{Area}(\gamma_A)}{4 G_N},
\end{equation}
where $\gamma_A$ is the co-dimension two minimal surface in the bulk that is anchored to the boundary of subsystem $A$ and homologous to it. For time-dependent scenarios, the proposal was extended by Hubeny, Rangamani, and Takayanagi (HRT) using extremal rather than minimal surfaces \cite{Hubeny:2007xt}. A key challenge in holography is understanding how bulk geometry encodes quantum correlations in the boundary theory. While MI captures bipartite correlations, it often fails to reflect the full structure of entanglement, particularly in deformed or interacting systems. The EWCS, conjectured to be dual to mixed state correlation measures such as EoP and EN, offers a refined probe of quantum correlations beyond holographic EE \cite{Takayanagi:2017knl,Nguyen:2017yqw}. 

In the holographic framework, particularly within the AdS/CFT correspondence, the thermofield double (TFD) state emerges as a critical construct for probing both quantum entanglement and chaotic dynamics in strongly coupled systems \cite{Shenker:2013pqa, Shenker:2013yza, Hartman:2013qma, Shenker:2014cwa, Roberts:2014isa, Fischler:2018kwt,Jahnke:2017iwi}. Serving as a purification of thermal states, the TFD enables a geometric realization of entangled boundary CFT theories through the dual description of a two-sided eternal AdS black hole \cite{Maldacena:2001kr}. This duality facilitates the investigation of non-local quantum correlations via refined observables like thermo mutual information (TMI), which is tailored to capture correlations beyond the reach of conventional MI \cite{Morrison:2012iz,Liu:2013iza,Leichenauer:2014nxa,Shenker:2014cwa, Jahnke:2018off}. Notably, in chaotic regimes, such correlations are highly sensitive to perturbations. An early-time perturbation can propagate rapidly and disrupt entanglement patterns—an effect referred to as scrambling. These phenomena are efficiently diagnosed through tools such as the shockwave geometry and pole-skipping techniques, revealing deep ties between horizon dynamics and operator growth. For further discussions on holographic chaos across various holographic backgrounds, we direct the reader to  \cite{Li:2024grs,Saha:2024bpt,Karan:2023hfk, Sil:2024vwf, Mahish:2022xjz, Sil:2020jhr,Baishya:2024gih, Chua:2025vig, Singh:2025kaz, Basu:2025exh,Lilani:2025wnd}.

Understanding the interplay between entanglement and chaos under deformations of the boundary theory is crucial for probing non-perturbative aspects of strongly coupled quantum systems. This includes the
generalizations of the Casini-Huerta-Myers (CHM) formalism \cite{Casini:2011kv} by Jensen and Bannon in \cite{Jensen:2013lxa}, that extended the method to boundary CFTs with codimension-$n$ defects, showing that the reduced density matrix maps to a thermal state on hyperbolic space. In the context of $\mathcal{N}_c=(2,0)$ Superconformal field theory (SCFT), \cite{Rodgers:2018mvq} employed this framework to study planar defects via holographic duals with probe $M2$/$M5$ branes in $AdS_7 \times S^4$. Symmetry-breaking deformations were analyzed by Carmi \cite{Carmi:2015dla} using the Ryu-Takayanagi prescription, revealing how EE responds to partial violations of translational and rotational invariance. Further work \cite{Carmi:2017ezk} explored the effects of scalar-induced bulk deformations on EE and complexity. Hung et al. \cite{Hung:2011ta} identified a logarithmic correction to EE from relevant operators, indicating the sensitivity of UV data to IR deformations.
Meanwhile, flavor backreaction effects were incorporated in $\mathcal{N}_c=4$ SYM by introducing $N_f$ D7-branes in AdS$_5$, with holographic EE computed to linear order in $N_f/N_c$ \cite{Kontoudi:2013rla}. The backreaction has also been modeled through smeared fundamental strings in an AdS black hole background, yielding a deformed AdS black hole geometry that is dual to a strongly coupled large $\mathcal{N}_c$ thermal field theory in presence of uniformly distributed heavy static fundamental quarks\cite{Chakrabortty:2011sp}. This setup was further utilized to study transport properties and entanglement measures \cite{Chakrabortty:2016xcb, Chakrabortty:2020ptb, Chakrabortty:2022kvq, Jain:2023xta}. Furthermore, in this context, the authors of \cite{Dey:2020yzl} construct an asymptotically AdS black hole solution incorporating higher‑derivative curvature corrections in the presence of a string cloud.

More recently, the inclusion of charges alongside gravitational backreaction has been considered in \cite{Pokhrel:2023plp, Dey:2023inw}. However, a comprehensive and systematic investigation into their combined influence on entanglement dynamics and chaotic behavior remains largely unexplored. The present study is motivated by a central question: how do additional physical parameters—namely, charge and backreaction alter the entanglement structure and signatures of quantum chaos in strongly coupled thermal field theories? These deformations introduce nontrivial dynamical scales into the bulk geometry, modifying the near-horizon and infrared structure of the dual spacetime. As a result, essential information-theoretic observables such as EE, MI, OTOC, and butterfly velocity are expected to exhibit qualitatively distinct behavior. Understanding these modifications is crucial for characterizing how charges and matter backreaction influence information scrambling, signal propagation and the emergence of effective geometries in holography.

In our study we find that the charge and backreaction significantly modify quantum correlations and enhance the scrambling. Holographically, they deform the bulk geometry and alter the connectivity of extremal surfaces leading to a faster disappearance of MI between distant regions. In contrast, the EWCS remains more robust, capturing deeper insights even when MI vanishes. 
Moreover, we characterize the chaos in terms of butterfly velocity and Lyapunov exponent using the shockwave analysis and study its dependence on charge and backreaction parameter. We find that an increase in both charge and the backreaction parameter leads to a decrease in the butterfly velocity $v_B$. We further compute the holographic TMI and observe that correlation between the boundary theories decreases with increasing the charge, suggesting that charge weakens these inter-boundary correlations. Finally, we find that when a shockwave is introduced, the disruption of holographic TMI becomes stronger in the presence of charge for finite non-zero backreaction.

The structure of this paper is as follows. In \cref{CbModel}, we introduce the background and setup for a charged backreacted AdS black hole. Next, \cref{sec:hee} focuses on the numerical study of HEE, while \cref{MI-EWCS} is dedicated to the analysis of MI and the EWCS. In \cref{sec:Shock-analysis}, we compute the butterfly velocity and Lyapunov exponent to characterize chaotic behavior. The behavior of TMI and its disruption due to perturbations are examined in \cref{htmi} and \cref{ShockandMI}, respectively. Finally, we summarize our findings in \cref{summary}.

\section{Background} \label{CbModel}

In this section we briefly review about the construction of the present setup as described in \cite{Pokhrel:2023plp}. Consider a \emph{charged $AdS_{d+1}$ black hole} in the presence of a \emph{string cloud} where the string cloud consists of open strings stretching from the AdS boundary to the black hole horizon. The holographic dual of this configuration corresponds to a large $\mathcal{N}_c$ strongly‑coupled field theory at finite temperature and finite chemical potential, with  uniform density of heavy static fundamental quarks (the quark‑cloud).

 To obtain the metric, we begin by defining the total action for the system which includes the Einstein-Maxwell action for a charged AdS black hole and the contribution from the string cloud. The total action is then given by
\begin{equation}
    \label{action_total}
    \mathcal{S} = -\frac{1}{16 \pi G_N^{d+1}} \int_M d^{d+1}x \, \sqrt{-g} \left( R - 2\Lambda - F_{\mu\nu} F^{\mu\nu} \right) + \mathcal{S}_{\text{SC}},
\end{equation}
where $G_N^{d+1}$ is the gravitational constant in $(d+1)$ dimensions, $g$ is the determinant of the metric $g_{\mu\nu}$, $R$ is the Ricci scalar, $\Lambda$ is the cosmological constant (negative for AdS) and $F_{\mu\nu}$ is the field strength tensor of the gauge field. The term $\mathcal{S}_{\text{SC}}$ represents the contribution from the string cloud, given by
\begin{equation}
    \label{string_action}
    \mathcal{S}_{\text{SC}} = -\frac{1}{2} \sum_i \mathcal{T}_i \int d^2 \xi \, \sqrt{h} \, h^{\alpha\beta} \, \partial_\alpha X^\mu \partial_\beta X^\nu g_{\mu\nu},
\end{equation}
where $\mathcal{T}_i$ is the tension of the $i^\text{th}$ string, $X^\mu(\xi)$ represents the embedding functions of the string worldsheet into spacetime and $h_{\alpha\beta}$ is the induced metric on the worldsheet with determinant $h$.

By extremizing the total action in \cref{action_total} with respect to the metric $g_{\mu\nu}$, one can obtain the Einstein–Maxwell equations with an additional source term arising from the string cloud \cite{Letelier:1979ej, Herscovich:2010vr}. This equation reads\footnote{Later on, we set $8 \pi G_N^{d+1} = 1$.}
\begin{equation}
    \label{equation_of_motion_wrt_g_munu}
    \frac{\delta \mathcal{S}}{\delta g_{\mu\nu}} = 0 \quad \Longrightarrow \quad R_{\mu \nu} - \frac{1}{2} g_{\mu \nu} R + \Lambda g_{\mu \nu} = T_{\mu \nu}^{EM} + 8 \pi G_{d+1} T_{\mu \nu}^{SC}.
\end{equation}
Here, the energy–momentum tensor for the electromagnetic field is given by
\begin{equation}
    \label{electromagnetic_stress_energy_tensor}
    T_{\mu \nu}^{EM} = 2 {F_\mu}^\lambda F_{\nu \lambda} - \frac{1}{2} g_{\mu \nu} F^2,
\end{equation}
and the energy–momentum tensor corresponding to the string cloud is
\begin{equation}
    \label{string_cloud_stress_energy_tensor}
    T^{(SC)\mu \nu} = - \sum_i \mathcal{T}_i \int d^2 \xi \, \frac{1}{\sqrt{-g}} \sqrt{h} \, h^{\alpha \beta} \, \partial_\alpha X^\mu \, \partial_\beta X^\nu \, \delta^{(d+1)}(x - X(\xi)),
\end{equation}
where the delta function localizes the string source in spacetime.
Assuming the strings are uniformly distributed over the $(d-1)$ spatial directions, the string cloud density $a(x)$ is defined as
\begin{equation}
    \label{string_density_equation}
    a(x) = T \sum_i \delta^{(d-1)}(x - X_i),
\end{equation}
where $T$ is the common string tension. Clearly, the density $a(x)$ depends on $x$. To treat it as a constant, we define the \emph{average density} $a$ as
\begin{equation}
    a = \frac{1}{V_{d-1}} \int dx^{d+1} \, a(x) = \frac{\mathcal{T} N}{V_{d-1}},
\end{equation}
where $V_{d-1}$ is the volume of the $(d-1)$-dimensional spatial section and $N$ is the total number of strings.
Furthermore, varying the action \eqref{action_total} with respect to the electromagnetic gauge field $A_\mu$ leads to the Maxwell equation
\begin{equation}
    \label{maxwell_equation_of_motion}
    \nabla_\alpha F^{\alpha\beta} = 0.
\end{equation}
The solution of the Einstein–Maxwell equations \cref{equation_of_motion_wrt_g_munu} in AdS spacetime, admits the following static, spherically symmetric metric ansatz:
\begin{equation}
    \label{metric_ansatz}
    ds^2 = -f_k(r) \, dt^2 + \frac{dr^2}{f_k(r)} + \frac{r^2}{R^2_{AdS}} \, d\Sigma_{k,d-1}^2,
\end{equation}
where $d\Sigma_{k,d-1}^2$ is the line element of a $(d-1)$-dimensional space of constant curvature $k$, given by
\begin{equation}
    d\Sigma_{k,d-1}^2 = \left\{
    \begin{array}{ll}
    R^2_{AdS} \, d\Omega^2_{d-1} & \text{for} \; k = +1, \\[6pt]
    \sum_{i=1}^{d-1} dx_i^2 & \text{for} \; k = 0, \\[6pt]
    R^2_{AdS} \, dH^2_{d-1} & \text{for} \; k = -1,
    \end{array}
    \right.
\end{equation}
where $d\Omega^2_{d-1}$ and $dH^2_{d-1}$ are the standard unit metrics on the $(d-1)$-sphere $S^{d-1}$ and hyperbolic space $H^{d-1}$, respectively. Here, $R_{AdS}$ is the AdS radius, which is related to the cosmological constant $\Lambda$ via the relation\footnote{The relation between the AdS radius $R_{AdS}$ and the cosmological constant $\Lambda$ in $(d+1)$ dimensions is given by
\begin{equation*}
    \Lambda = - \frac{d(d-1)}{2 R^2_{AdS}}.
\end{equation*}
}
Also, one can solve the Maxwell equation given in \cref{maxwell_equation_of_motion} by assuming a static electric field with the potential 
\[
A_{\mu} = (A_t(r), 0, 0, 0),
\]
which leads to the following solution\footnote{Here, we consider the static electric potential in the black hole background.}
\begin{equation}
    \label{maxwell_equation_of_motion_solution}
    F_{rt} = \sqrt{\frac{(d-1)(d-2)}{2}} \, \frac{q}{r^{d-1}},
\end{equation}
where \(q\) is a charge parameter related to the total charge \(Q\) of the black hole and given by
\begin{equation}
    \label{charge_Q_value}
    Q = \sqrt{2 (d-1)(d-2)} \, q \, \omega_{d-1}.
\end{equation}
Here \(\omega_{d-1}\) denotes the volume of the unit \((d-1)\)-dimensional spatial hypersurface.
The chemical potential \(\Phi\) induced by the black hole charge can be expressed as
\begin{equation}
    \label{electrostatic_potential_equation}
    \Phi = \sqrt{\frac{(d-1)}{2(d-2)}} \, \frac{q}{r_+^{d-2}},
\end{equation}
where \(r_+\) is the outer horizon radius of the black hole. For the string cloud energy-momentum tensor \(T^{(SC)}_{\mu\nu}\), using the static gauge condition, i.e.,
\[
X^t = \xi^0, \quad X^r = \xi^1, \quad X^\mu = y^\mu \quad \text{for} \quad \mu \ne t, \, r,
\]
the non-vanishing components are
\begin{equation}
    \label{string_cloud_non_vanishing_components}
    T^{(SC)tt} = -\frac{a g^{tt}}{r^{d-1}}, \quad T^{(SC)rr} = -\frac{a g^{rr}}{r^{d-1}},
\end{equation}
or equivalently:
\begin{equation}
    {T^{(SC)}}^t_t = {T^{(SC)}}^r_r = -\frac{a}{r^{d-1}},
\end{equation}
with \(a > 0\) representing the string cloud density parameter \cite{Chakrabortty:2011sp}. This ensures that the energy conditions are satisfied.

On solving the Einstein-Maxwell equations described in \cref{equation_of_motion_wrt_g_munu} with the energy-momentum tensors discussed earlier and the metric ansatz in \cref{metric_ansatz}, we obtain the solution for the metric function \(f_k(r)\) as
\begin{equation}
    \label{metric_solution_V_with_K}
    f_k(r) = k + \frac{r^2}{R^2_{AdS}} - \frac{2m}{r^{d-2}} + \frac{q^2}{r^{2d-4}} - \frac{2 a R^{d-3}_{AdS}}{(d-1) r^{d-3}},
\end{equation}
where \(m\) is an integration constant associated with the black hole mass.
For simplicity, we focus on the case \(k=0\), corresponding to an AdS black hole with a flat boundary. The metric function then reduces to
\begin{equation}
    \label{f(r)}
    f(r) = \frac{r^2}{R^2_{AdS}} - \frac{2m}{r^{d-2}} + \frac{q^2}{r^{2d-4}} - \frac{2 a R^{d-3}_{AdS}}{(d-1) r^{d-3}}.
\end{equation}
Further, we can write the metric of \cref{metric_ansatz} in terms of Poincaré coordinates defined by \(r = \frac{R^2_{AdS}}{z}\), where \(R_{AdS}\) is the AdS radius. This yields\footnote{We omit the subscript \(k\) in the blackening function after fixing \(k=0\).}
\begin{equation}
    \label{pmetric}
    ds^2 = \frac{R^2_{AdS}}{z^2} \left(-f(z)dt^2 + \frac{dz^2}{f(z)} + d\vec{x}^2_{d-1} \right),
\end{equation}
where the blackening function \(f(z)\) is given by
\begin{equation}
    \label{fz}
    f(z) = 1 - \rho \left( \frac{z}{z_h} \right)^{d-1} + (\rho - \sigma - 1) \left( \frac{z}{z_h} \right)^{d} + \sigma \left( \frac{z}{z_h} \right)^{2d-2}.
\end{equation}
Here, we introduced the dimensionless parameter \(b\) for the string cloud density by scaling \(a\) as \(b = aR^{d-1}\). The dimensionless parameters \(\rho\) and \(\sigma\) are defined as
\begin{equation}
    \label{rs}
    \rho = \frac{2bz_h^{d-1}}{(d-1)R^{d-1}_{AdS}}, \qquad \sigma = \frac{q^2 z_h^{2d-2}}{R^{4d-6}_{AdS}},
\end{equation}
where \(z_h\) is the location of the horizon, and \(\rho\) and \(\sigma\) encode the effects of the string cloud density and the black hole charge, respectively.

The temperature of the black hole is obtained by demanding regularity of the Euclidean metric at the horizon. The Hawking temperature is given by
\begin{equation}
    \label{hawking_temperature_formula}
    T = -\frac{1}{4\pi} \left[ \frac{df(z)}{dz} \right]_{z=z_h}.
\end{equation}
Using the explicit form of \(f(z)\) in \cref{fz}, we obtain
\begin{equation}
    T = \frac{1}{4\pi z_h} \left( d - \rho - (d - 2)\sigma \right).
\end{equation}
This result can be derived by following the standard procedure for computing the surface gravity at the black hole horizon.
The positivity of temperature $(T\geq0)$ imposes the following inequality:
\begin{equation}
    d \geq \frac{\rho - 2\sigma}{1-\sigma },
\end{equation}
which provides a bound on the parameters \(\rho\) and \(\sigma\) for a well-defined thermodynamic system. To understand the other thermodynamical properties such as entropy, free energy, Landau function and the black hole phases, refer to \cite{Pokhrel:2023plp}.

\section{Holographic Entanglement Entropy} \label{sec:hee}

\begin{figure}[H]
\centering
\includegraphics[width=.40\linewidth]{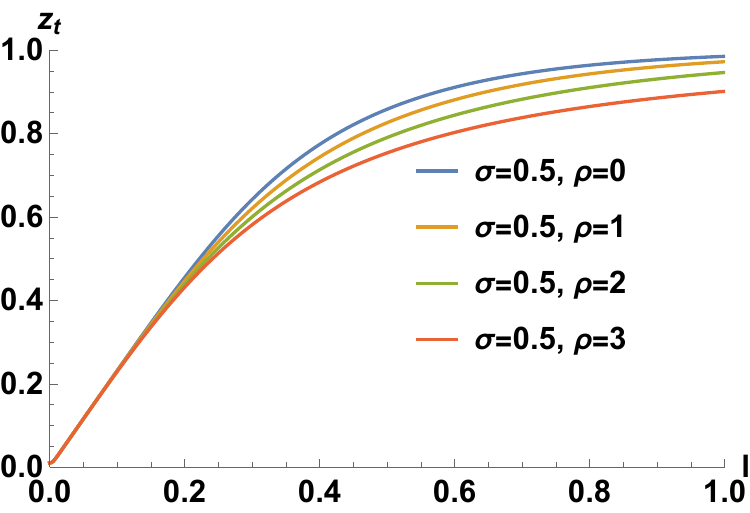}
 \includegraphics[width=.40\linewidth]{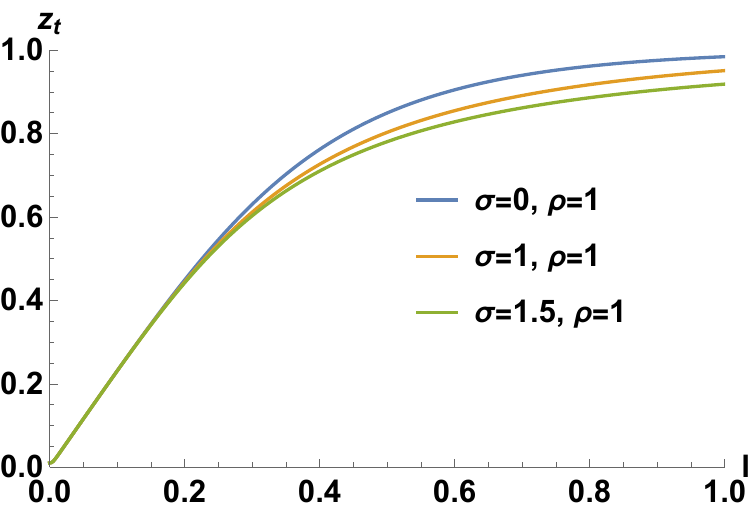}
\caption{\textbf{Left:} Variation of RT-surface for constant $\sigma$ and different $\rho$. \textbf{Right:} Variation of RT-surface for fix $\rho$ and different values of $\sigma$.}
\label{ztl}
\end{figure}

In this section we compute the holographic entanglement entropy of deformed $AdS_{d+1}$ charged black hole discussed in \cref{CbModel}. The subsystem  $A$ is described by a $(d-1)$-dimensional spatial long rectangular strip of widths $l$ and $L$ as

\begin{equation}
  x\equiv x^1\in \left[-\frac{l}{2},\frac{l}{2}\right]~,~x^i \in  \left[-\frac{L}{2},\frac{L}{2}\right]~,i=2,...,d-2,
\end{equation}
where $L\gg l$ is taken to preserve invariance of the subsystem in $x^i$ directions. The HEE for subsystem $A$ can be obtained using the RT formula in the background geometry as described in \cref{pmetric} which is given by 

\begin{equation}\label{RT-area}
\begin{aligned}
        S_A&=\frac{1}{4 G^{d+1}_N} 
   \left(\prod _{i=2}^{d-2} \int_{-L/2}^{L/2} dx^i\right)
    \left[  2\int_{0} ^{z_t}dz \sqrt{ g_{in}^{A} } \right]\\
    &=\frac{
    L^{d-2}R^{d-1}_{AdS}}{2G^{d+1}_N} 
    \int_{0}^{z_t}\frac{dz}{z^{d-1}\sqrt{f(z)}\sqrt{1-\left(\frac{z}{z_t} \right)^{2d-2}}},
    \end{aligned}
\end{equation}
where $g_{in}^{A}$ is the induced metric on the minimal surface corresponding to subsystem $A$ and $z=z_t$ is turning point of the minimal surface satisfying $z'(x)=0$. The relation between the width of the subsystem $A$ and $z_t$ is given by the following relation
\begin{equation}\label{turning-point}
l/2=1/z_t^{d-1}\int_{0}^{z_t} z^{d-1}\left[f(z)\left(1-(z/z_t)^{2d-2}\right)\right]^{-1/2} dz.
\end{equation}
Since solving the integral in \cref{turning-point} is particularly difficult for arbitrary dimensions $d$, we use a numerical technique to study the behavior of the RT surface in $d=4$ with respect to the width $l$, for particular $\rho$ and the $\sigma$ as shown in \cref{ztl}.

From \cref{ztl} (left) one can see that for a finite constant $\sigma$ as the value of width $l$ increases the RT surface travels more deeper into the bulk. However, by increasing the string density $\rho$ makes it slower. Similar behavior of RT surface can be seen with respect to the charge parameter $\sigma$ for constant $\rho$.  
Using the expression of turning point \cref{turning-point} and RT formula \cref{RT-area} we can analyse the HEE  with respect to width as well as $\rho$ and $\sigma$ parameters.
\begin{figure}[H]
\centering
 \includegraphics[width=.40\linewidth]{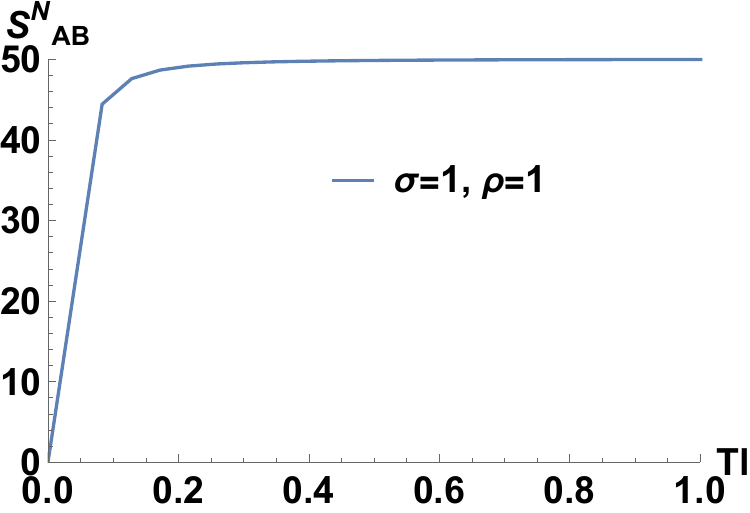}
 \includegraphics[width=.40\linewidth]{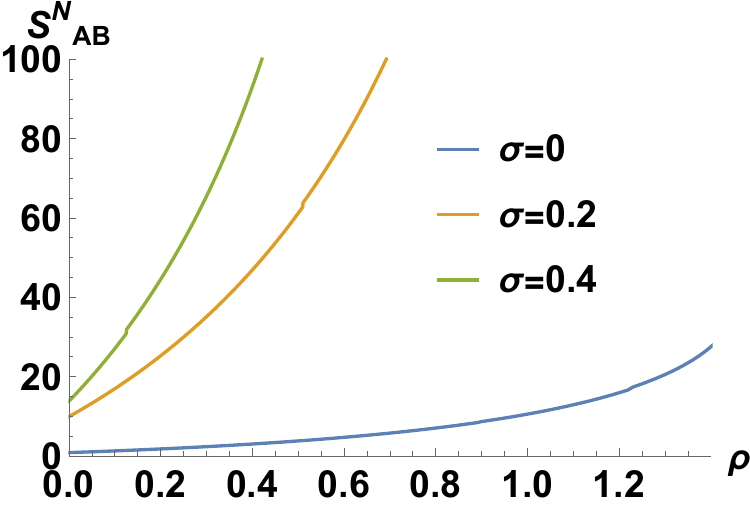}
\caption{\textbf{\textit{Left:}}
        normalized HEE (\( S^N_{AB}\)) as a function of width \( l \) at fixed temperature \( T = 1 \), backreaction \( \rho = 1 \), and charge \( \sigma = 1 \). 
        \textbf{\textit{Right:}} HEE as a function of \( \rho \) for various values of \( \sigma \). HEE increases with both \( \rho \) and \( \sigma \), reflecting the enhanced contribution to entanglement entropy.}
\label{Svslfg}
\end{figure}
The behavior of the \( S^N_{AB}\) is plotted in \cref{Svslfg}. This analysis reveals several important features about the entanglement structure of strongly coupled field theory influenced by charge and backreaction:

\begin{itemize}
    \item From \cref{Svslfg} (Left) we observe that HEE increases monotonically with width \( l \), consistent with the expectation that larger subsystems capture more degrees of freedom leading to higher entanglement entropy. The growth rate of HEE diminishes for larger \( l \), indicating a saturation tendency. For entanglement entropy, the thermal correlations dominate at large subsystem size at fixed temperature and the contribution from quantum entanglement becomes subleading. 
    
    \item  The presence of backreaction (\( \rho = 1 \)) and charge (\( \sigma = 1 \)) introduces additional bulk degrees of freedom, modifying the geometry and influence the dual field theory observables such as correlation functions and entanglement structure. In certain cases, these effects can be interpreted as modifying the effective number of degrees of freedom, though this does not necessarily correspond to a simple increase in the central charge. In the bulk, the extremal surface anchored to the boundary grows deeper into the geometry as \( l \) increases. At finite temperature and in the presence of charge, the surface tends to probe regions near the black hole horizon, where contributions from thermal and charged matter fields enhance the area and thus the HEE.
    \item  In \cref{Svslfg} (Right) for all values of \( \sigma \), the HEE increases with increasing backreaction parameter \( \rho \). This reflects the fact that stronger backreaction introduces more matter content in the bulk geometry, which corresponds to a larger number of active degrees of freedom in the dual boundary field theory, thereby enhancing the total entanglement entropy. The increasing behavior of HEE w.r.t $\rho$ is also reported in \cite{Chakrabortty:2020ptb} for zero charge case i.e. for $\sigma=0$.
    \item At fixed \( \rho \), the HEE increases with the charge parameter \( \sigma \). This indicates that charge also contributes additional degrees of freedom that enhance the entanglement structure in the boundary theory. 
    In the holographic dual, the extremal surface area increases with both \( \rho \) and \( \sigma \), due to more significant warping of the bulk geometry and deeper penetration into regions near the horizon. This geometric stretching accounts for the observed rise in HEE.
\end{itemize}

We have demonstrated that the HEE exhibits an increasing trend with both the backreaction and the charge parameter. A similar behavior of HEE has been observed in backreacted model \cite{Chakrabortty:2020ptb}, where an increase in quark content leads to an enhancement in external degrees of freedom, thereby resulting in greater entanglement within the system. In the present analysis, we further establish that an increase in the charge also leads to an enhancement of HEE. This can be attributed to the fact that adding charge introduces additional degrees of freedom into the system, thus amplifying the EE. However, given the limitations of HEE as a measure in mixed states, we proceed in the next section to examine the influence of these parameters on holographic mutual information and EWCS which serves as more appropriate entanglement measures in the case of mixed state.

\section{Holographic Mutual Information and EWCS}\label{MI-EWCS}

To analyze the influence of the relevant parameters on mixed-state entanglement, we focus on two prominent measures: the holographic mutual information (HMI) and the entanglement wedge cross section (EWCS). We begin by considering the mutual information (MI), which captures the total correlation comprising both quantum and classical contributions between two subsystems. Given a bipartite decomposition of the system into regions $A$ and $B$, the mutual information is defined as
\begin{equation}\label{midf}
I(A, B) = S_A + S_B - S_{A \cup B},
\end{equation}
where $S_A $, $S_B $ and $S_{A \cup B}$ denote the von Neumann entanglement entropies associated with the subsystems $A$, $B$, and the union $A \cup B$, respectively. This quantity serves as a diagnostic tool for identifying correlations and phase transitions in quantum systems, particularly in holographic and strongly correlated regimes. Consider two disjoint rectangular subsystems $A$ and $B$ of length $l$, separated by a distance $D$ along the $x^1$ direction and extended by $L$ along the transverse directions. The HEE corresponding to a rectangular subsystem of width $l$ is given by \cref{RT-area} while, for combined system $A \cup B$, there are two possible RT surfaces. When the separation between the subsystems is large ($D/l \gg 1$), the minimal surface becomes disconnected, giving $S_{A \cup B} = S_A + S_B$ and thus $I(A, B) = 0$, indicating disentanglement. In contrast, for small separations ($D/l \ll 1$), the connected surface dominates, yielding
\begin{equation}\label{sld}
S_{A \cup B} = S(2l + D) + S(D),
\end{equation}
and gives the nonzero $I(A, B)$.

\begin{figure}[H]
\centering
 \includegraphics[width=.50\linewidth]{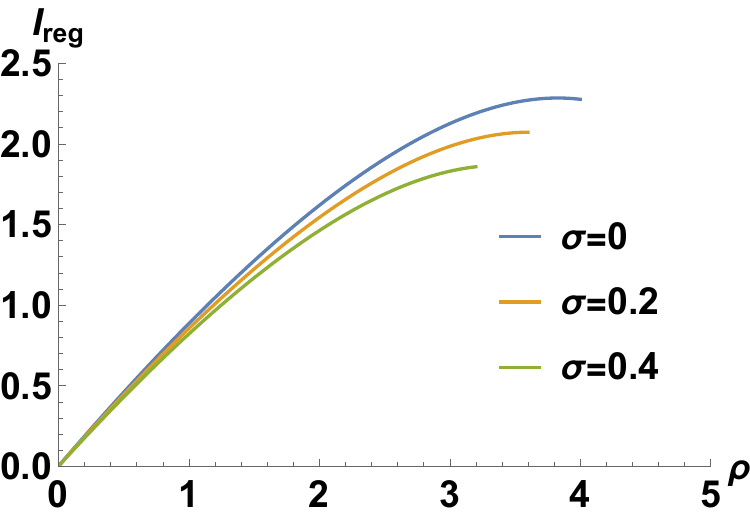}
 \caption{
        Regularized \( I(A:B) \) as a function of backreaction \( \rho \), for different values of \( \sigma \). The widths of subsystems are kept fix by fixing the corresponding turning points $z_t(D)=0.01, z_t(l)=0.4$ and $z_t(2l+D)=0.7$. The mutual information exhibits an increasing nature with \( \rho \). Higher charge consistently reduces MI, indicating weakened inter-subsystem correlations due to additional charged degrees of freedom in the strongly coupled field theory.}
\label{Ivsrho}
\end{figure}

Since we are only interested in the nonzero MI therefore we will only consider the case where the separation between the subsystem is small i.e $(D/l\ll 1)$. From \cref{midf} and \cref{sld} we can write the MI in the following form 
\begin{equation}\label{mild}
    I(A,B)=2S(l)-S(2l+D)-S(D).
\end{equation}
Now using \cref{RT-area} in above \cref{mild} one can write the integral form of $I(A,B)$ as
\begin{equation}\label{eq:mi}
\begin{aligned}
     I(A,B)= & \frac{ L^{d-2}R^{d-1}_{AdS}}{2G_N^{d+1}} 
    \biggl[2
    \int_{0}^{z_t(l)}\frac{dz}{z^{d-1}\sqrt{f(z)}\sqrt{1-\left(\frac{z}{z_t} \right)^{2d-2}}} \\
    &-\int_{0}^{z_t(2l+D)}\frac{dz}{z^{d-1}\sqrt{f(z)}\sqrt{1-\left(\frac{z}{z_t} \right)^{2d-2}}}-\int_{0}^{z_t(D)}\frac{dz}{z^{d-1}\sqrt{f(z)}\sqrt{1-\left(\frac{z}{z_t} \right)^{2d-2}}}
    \biggr].
\end{aligned}
\end{equation}
Similar to the case of HEE, the HMI integral in \cref{eq:mi} is analytically intractable. Consequently, we resort to a numerical evaluation of HMI by fixing the relevant parameters. In \cref{Ivsrho}, we present the behavior of HMI as a function of the backreaction parameter $\rho$ for several representative values of the parameter $\sigma$.

In \cref{Ivsrho} we plot the behavior of the holographic mutual information as a function of the backreaction parameter \( \rho \), for different values of the charge parameter \( \sigma \). The following observations can be made:

\begin{itemize}
    \item As \( \rho \) increases the MI increases as well, indicating that the backreaction enhances the correlations between the two subsystems. This suggest that even in presence of charge the backreaction act as the enhancer for the correlations but the rate of increment becomes slower as we increase the charge. Also due to the bond between $\rho$ and $\sigma$, the MI curves ends at particular values of $\rho$. 
\item  At fixed backreaction \( \rho \) the holographic MI decreases as the charge parameter \( \sigma \) increases. This reduction reflects a weakening of long-range correlations in the dual field theory.
\end{itemize}

 Next, we numerically investigate the influence of the charge and the backreaction parameter on the EWCS which is defined as the minimal area of surface that partitions the entanglement wedge of \(A\cup B\) into two parts associated with \(A\) and \(B\) and given by \cite{Takayanagi:2017knl,Nguyen:2017yqw}
\[
E_W(A:B) = \frac{\mathrm{Area}(\Sigma_{AB})}{4G_N^{d+1}},
\]
where \(\Sigma_{AB}\) is the minimal cross-sectional area of the entanglement wedge associated with the boundary regions \(A\) and \(B\).
The EWCS is conjectured to be dual to the \textit{entanglement of purification} for holographic states and bounds the MI from above \cite{Takayanagi:2017knl,Nguyen:2017yqw}. In particular, for MI the following inequalities are satisfied:
\begin{equation}
\frac{1}{2} I(A:B) \leq E_W(A:B) \leq \min \left\{ S_A, S_B \right\}.
\end{equation}

\begin{figure}[H]
\centering
 \includegraphics[width=.60\linewidth]{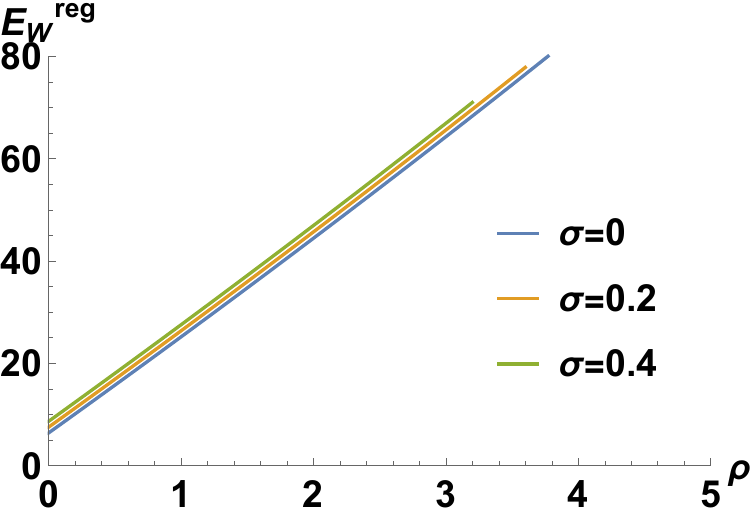}
 \caption{Regularized $E_W$ as a function of \( \rho \) for various values of \( \sigma \). EWCS increases monotonically with both \( \rho \) and \( \sigma \), indicating enhanced entanglement due to additional degrees of freedom introduced by backreaction and charge. This reflects a richer entanglement structure in the dual strongly coupled field theory.}
\label{ewfg}
\end{figure}

Following the prescription outlined in \cite{Takayanagi:2017knl}, the integral representation of the EWCS is given by
\begin{equation}
   E_W(A:B)= \frac{
    L^{d-2}R^{d-1}_{AdS}}{4G_N^{d+1}} 
    \left[\int_{z_t(D)}^{z_t(2l+D)}\frac{dz}{z^{d-1}\sqrt{f(z)}}
    \right].
\label{eq:ewcs}
\end{equation}
 We plot the above expression in \cref{eq:ewcs} for specific values of parameters as shown in \cref{ewfg}.

The above figure shows the behavior of the EWCS as a function of the backreaction \( \rho \), with each curve corresponding to a different value of the charge parameter \( \sigma \). The following conclusions can be drawn:

\begin{itemize}
    \item For all values of \( \sigma \), EWCS increases monotonically with \( \rho \). This indicates that stronger backreaction, which encodes additional bulk matter or stringy corrections, leads to more robust bulk entanglement between the boundary subsystems. This is consistent with an enhanced capacity for mixed-state entanglement in the dual strongly coupled field theory as reported in \cite{Chakrabortty:2020ptb}.

    \item  At any fixed value of \( \rho \), the EWCS increases with increasing  \( \sigma \). Physically, this suggests that charge enhances the total correlation content shared between the subsystems, possibly due to the presence of new degrees of freedom that enrich the internal structure of correlations.
    \item In the holographic dual, the EWCS is associated with the minimal cross-sectional area of the entanglement wedge. As \( \rho \) and \( \sigma \) increase, the bulk geometry becomes richer allowing the wedge cross section to grow. This growth geometrically encodes enhanced entanglement in the boundary theory.
\end{itemize}

\begin{figure}[H]
\centering
 \includegraphics[width=.50\linewidth]{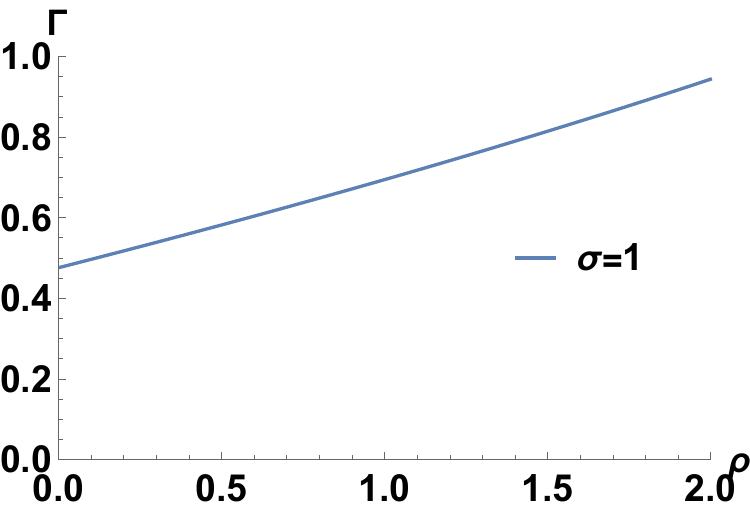}
\caption{The bound between $E_w$ and $I/2$ for $\sigma=1$.}
\label{EIbound}
\end{figure}

The EWCS provides an upper bound on the mutual information, satisfying the inequality $E_W \geq \frac{I}{2}$. This bound is explicitly verified in \cref{EIbound} for a range of values of $\rho$ at a fixed charge parameter $\sigma = 1$. Furthermore, we have confirmed that the inequality holds for all values of $\sigma$.
In \cref{EIbound} we examine the known inequality \( \Gamma=E_W - \frac{1}{2} I(A:B) \geq 0\) as a function of the backreaction parameter \( \rho \), for  \( \sigma =1\). We observe that across all values of \( \rho \) and \( \sigma \), the plotted quantity \( \Gamma \) remains non-negative, consistent with the inequality \( E_W \geq \frac{1}{2} I(A:B) \). This reaffirms the holographic expectation that entanglement wedge cross section captures a deeper quantum correlation structure than mutual information alone.

So far, we have examined the influence of charge and backreaction parameters on the entanglement structure of the theory discussed in \cref{CbModel}, focusing on HEE, HMI and EWCS. In the following section, we shift our attention to the chaotic aspects of the theory. Specifically, we investigate how the presence of charge along with backreaction modifies its chaotic behavior. We compute the butterfly velocity and the Lyapunov exponent using holographic technique called shockwave analysis.

\section{Butterfly Velocity and Lyapunov Exponent}
\label{sec:Shock-analysis}
It is well understood that the thermofield double (TFD) state has a holographic dual description that plays a crucial role in the study of  holographic chaos and correlations \cite{Shenker:2013pqa,Hartman:2013qma, Shenker:2013yza,Stanford:2014jda, Morrison:2012iz, Jahnke:2017iwi,Jahnke:2018off}. More precisely, the eternal AdS black hole geometry corresponds to the TFD state defined in a doubled Hilbert space comprising two identical, non-interacting copies of the boundary conformal field theory (CFT)\cite{Maldacena:2001kr, Israel:1976ur}.
In this section, we investigate the chaotic behavior of a  backreacted strongly coupled large-$\mathcal{N}_c$ thermal field theory with finite chemical potential that is dual to a charged AdS black hole with string cloud. We compute the butterfly velocity ($v_B$) and the Lyapunov exponent ($\lambda_L$), the fundamental quantities describing the quantum chaotic nature of the theory using the shockwave method that requires the eternal black hole description \cite{Shenker:2013pqa}.

\begin{figure}[H]
\centering
 \includegraphics[width=.30\linewidth]{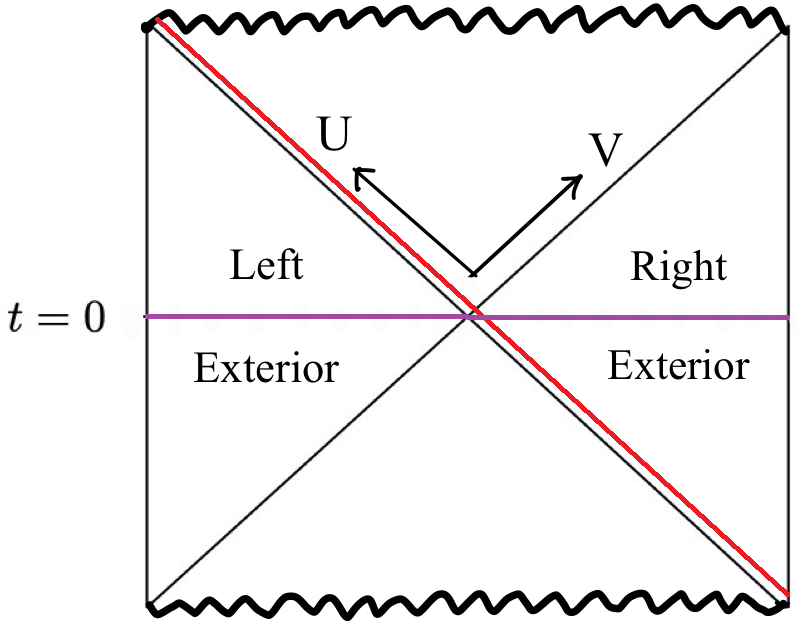}
 \hspace{6mm}
 \includegraphics[width=.30\linewidth]{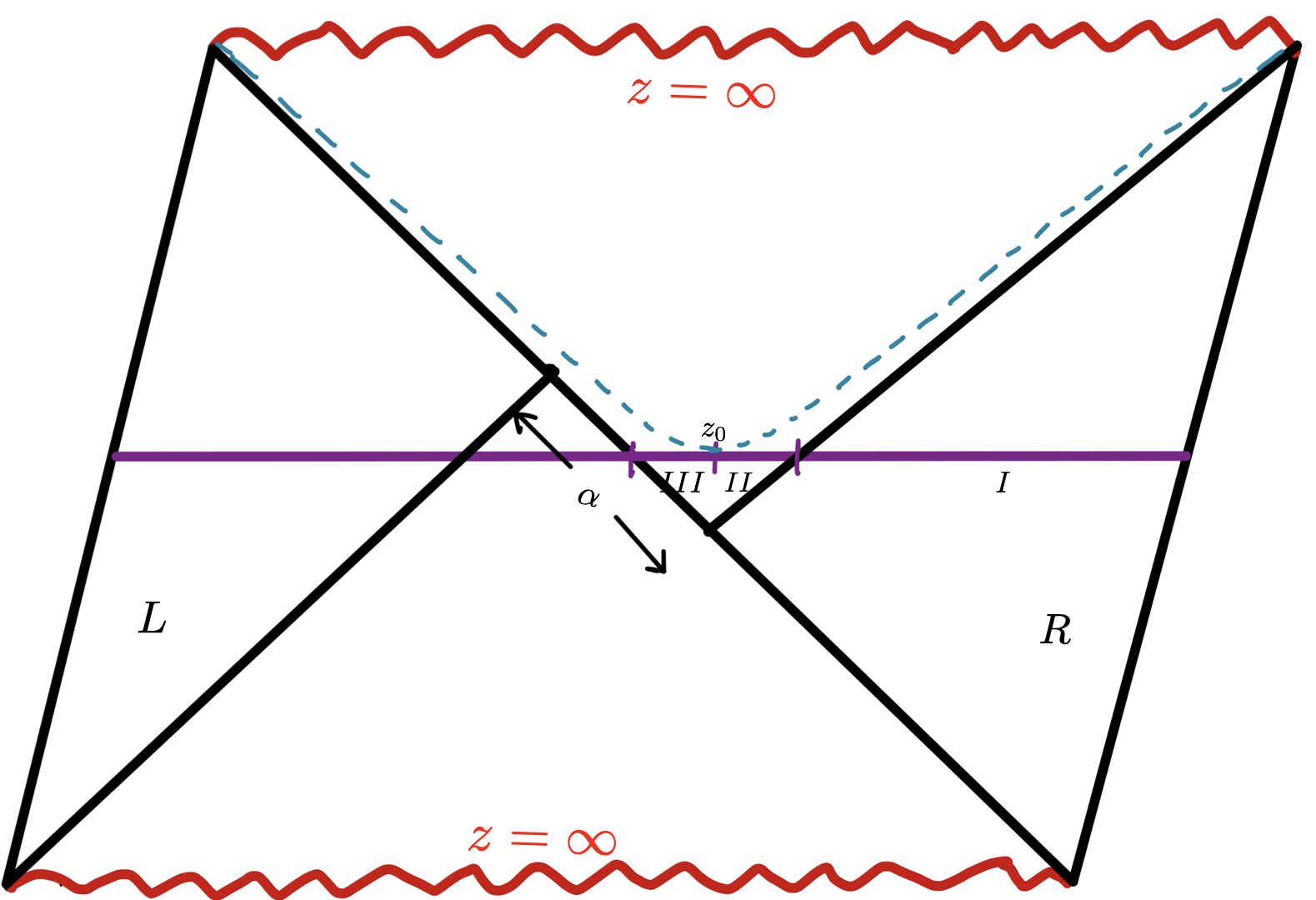}
\caption{{\bf {\textit{Left:}}} Penrose diagram of the eternal, backreacted charged AdS black hole, with the shock wave indicated by the red line. The purple curve represents the HRT surface traversing the bifurcation point of the black hole horizon.
 {\bf {\textit{Right:}}} Deformed Penrose diagram illustrating the effect of the shock wave on the black hole geometry. The horizons are displaced, allowing the HRT surface to probe deeper into the black hole interior. Adapted from\cite{Chakrabortty:2022kvq}.} 
\label{eternalBH}
\end{figure}

 Consider two identical, non-interacting boundary quantum field theories, each describing a large-$\mathcal{N}_c$, strongly coupled system at finite temperature and finite chemical potential, further deformed by the quark cloud. The TFD state constructed from these two copies defines an entangled pure state that holographically corresponds to an eternal, two-sided, charged AdS black hole geometry deformed by the string cloud.
An infinitesimal in-falling perturbation, modeled as a null pulse of energy with $\mathcal{E}_0$ at the boundary, is introduced from the right boundary of the backreacted charged eternal AdS black hole. This perturbation is depicted by the red null trajectory in \cref{eternalBH} (left). 
This perturbation originates from the asymptotic boundary at an early time and subsequently propagates towards the horizon at $U=0$. As it reaches the horizon at $t=0$, the perturbation undergoes an exponential blue shift, transforming into a shockwave that alters the bulk spacetime into a shockwave geometry as shown in \cref{eternalBH} (right).

The unperturbed geometry can be expressed in terms of Kruskal-Szekeres coordinates $(U,V)$ that covers the extended space-time of metric in \cref{pmetric} as follows
\begin{equation}\label{unperturbed-metric}
    ds^2 = 2\mathcal{B}(U,V)dUdV + g_{xx}(U,V)d\vec{x}_{d-1}^2,
\end{equation}
where the function $\mathcal{B}(U,V)$ is defined by
\begin{equation}
    \mathcal{B}(U,V) = \frac{\beta^2}{8\pi^2} \frac{|g_{tt}(U,V)|}{UV}.
\end{equation}
Here $\beta$ is the inverse temperature and $g_{tt}(U,V)$ is the temporal component of the metric. The coordinate transformations between coordinates $(U,V)$ and Poincare coordinates $(t,z)$ for the right half of eternal black hole are
\begin{equation}
    U = - e^{-\frac{2\pi}{\beta} (t - z_*)}, \quad V =  e^{\frac{2\pi}{\beta} (t + z_*)},
\end{equation}
with $t$ being the time and $z_*$ is the tortoise coordinate defined by
\begin{equation}
    z_* = - \int_0^z \sqrt{\frac{1}{f(z)^2}} dz.
\end{equation}
The general form of energy-momentum tensor corresponding to the unperturbed (without shock) metric \cref{unperturbed-metric} is
\begin{equation}
    T^{\text{matter}}_0 = 2 T_{{U}{V}} dU dV + T_{{U}{U}} dU^2 + T_{{V}{V}} dV^2 + T_{ij} dx^i dx^j,
\end{equation}
with $T_{\mu \nu} \equiv T_{\mu \nu} (U, V, \vec{x})$.
We consider a massless perturbation (pulse) traveling at the speed of light along the \( U \)-direction from the right asymptotic boundary, localized at \( V = 0 \), and aim to study its impact on the geometry described by \cref{unperturbed-metric}. Following the approach discussed in \cite{DRAY1985173,Sfetsos:1994xa}, we adopt the ansatz that the spacetime is given by \cref{unperturbed-metric} for \( V < 0 \), and for \( V > 0 \), the same metric applies but with the coordinate shifted as \( U \to U + \alpha \) while keeping all other coordinates unchanged. Here \( \alpha \) is the function to be determined and represents the strength of the perturbation or shockwave. Accordingly, the complete form of the spacetime metric and the energy-momentum tensor becomes:
\begin{equation}\label{shockwave-metric}
    ds^2 = 2 \mathcal{B}(\hat{U},\hat{V}) d\hat{U} d\hat{V} - 2 \mathcal{B}(\hat{U},\hat{V}) \alpha \delta(\hat{V}) d\hat{V}^2 + g_{xx}(\hat{U},\hat{V}) d\vec{\hat{x}}^2.
\end{equation}
and
\begin{equation}\label{semt}
\begin{split}
    T^{\text{matter}}
=& 2\left[T_{{U}{V}}-T_{{U}{U}}\alpha\delta(\hat{V})\right]d\hat{U}d\hat{V} +T_{{U}{U}}d\hat{U}^2
+T_{ij}d{x^i}d{x^j}
\\&+\left[T_{{V}{V}}-2T_{{U}{V}}\alpha\delta(\hat{V})+T_{{U}{U}}{\alpha}^2 
\delta(\hat{V})^2\right]d\hat{V}^2 ,
\end{split}
\end{equation}
where $\hat{U} = U + \theta(V) \alpha(t,\vec{x}),  \hat{V} = V$,
 $\hat{x}^i=x^i$ and also the function $\theta(V)$ ensures the changes in $V>0$ part only.
The energy-momentum tensor corresponding to the shockwave is given by
\begin{equation}\label{shockwave-EMT}
    T^S = \mathcal{E}_0 e^{\frac{2\pi }{\beta}t} \delta(\hat{V}) \delta (x^i) d\hat{V}^2,
\end{equation}
where $\mathcal{E}_0$ is the initial energy of the pulse and $e^{\frac{2\pi }{\beta}t}$ represents the fact that the energy of pulse is exponentially increasing with time $t$, that accounts for the blue shift. The form of the shift function $\alpha$ can be obtained by solving the Einstein field equations with \cref{shockwave-metric}, \cref{semt} and \cref{shockwave-EMT}
\begin{equation}
    R_{\mu \nu} - \frac{1}{2} g_{\mu \nu} R + \Lambda g_{\mu \nu} = 8\pi G_N \left( T^{\text{matter}}_{\mu \nu} + T^S_{\mu \nu} \right).
\end{equation}
By introducing a parameter $\epsilon$ such that $\alpha \to \epsilon\alpha $ and $T^S_{\mu \nu}\to \epsilon T^S_{\mu \nu}$ so that $\epsilon \to 0$ completely remove the effect of shock. Finally, we obtain the following differential equation

\begin{equation}\label{alpha-equation}
    \left( \partial_x^2 - M^2 \right) \alpha(t,\vec{x}) = \frac{g_{xx}(z_h)}{A(z_h)} 8\pi E_0 e^{\frac{2\pi}{\beta}t} \delta(x),
\end{equation}
where
\begin{equation}
    M^2 = \frac{(d-1)z_h}{2A(z_h)}\partial_{z}g_{xx}(z_h).
\end{equation}
The unique solution consistent with the delta source at $x=0$ and exponential decay at spatial infinity is given by
\begin{equation}\label{alpha-solution}
    \alpha(t,\vec{x}) \sim e^{\left[\frac{2\pi}{\beta}(t - t_*) - Mx \right]},
\end{equation}
where $t_*$ represents the scrambling time.

In the study of quantum many-body dynamics, especially within strongly coupled systems, sensitivity to initial conditions is a defining feature of quantum chaos. A standard diagnostic of such chaotic behavior is the \emph{butterfly effect}, which is characterized by the \emph{butterfly velocity} and the \emph{Lyapunov exponent}. These quantities are typically extracted from the growth of the \emph{out-of-time-order correlator} (OTOC) which serves as a quantitative probe of information scrambling in quantum systems.
 For two generic Hermitian operators \( W(x,t) \) and \( V(0) \), the thermal OTOC is defined by the thermal expectation value as
\begin{equation}
    C(t, x) = -\langle [ W(x, t), V(0)]^2 \rangle_{\beta},
\end{equation}
where  \( x \) is the spatial separation between the operators. This quantity measures the extent to which an initially local perturbation affects distant degrees of freedom as the system evolves in time.

In the framework of large-$\mathcal{N}_c$ gauge theories or their holographic duals with a semiclassical gravity description, it is well established that, at early times following the onset of scrambling, the out-of-time-ordered correlator (OTOC) exhibits exponential growth \cite{Shenker:2013pqa,Shenker:2013yza, Roberts:2014isa}. Specifically, the commutator squared takes the form
\begin{equation}
    C(t, x) \approx \frac{K}{N_c^2} \, e^{\lambda_L \left( t - t_* - \frac{x}{v_B} \right)},
\end{equation}
where $K$ is an $\mathcal{O}(1)$ constant, $\lambda_L$ denotes the Lyapunov exponent that governs the rate of exponential growth and $v_B$ is the butterfly velocity, which characterizes the spatial propagation of perturbations. The parameter $t_*$ represents the scrambling time, given by
\begin{equation}
    t_* \sim \frac{\beta}{2\pi} \log N_c^2,
\end{equation}
with $\beta$ being the inverse temperature of the system. This behavior captures the onset of quantum chaos in strongly coupled many-body systems with holographic duals.

\begin{figure}[H]
\centering
 \includegraphics[width=.50\linewidth]{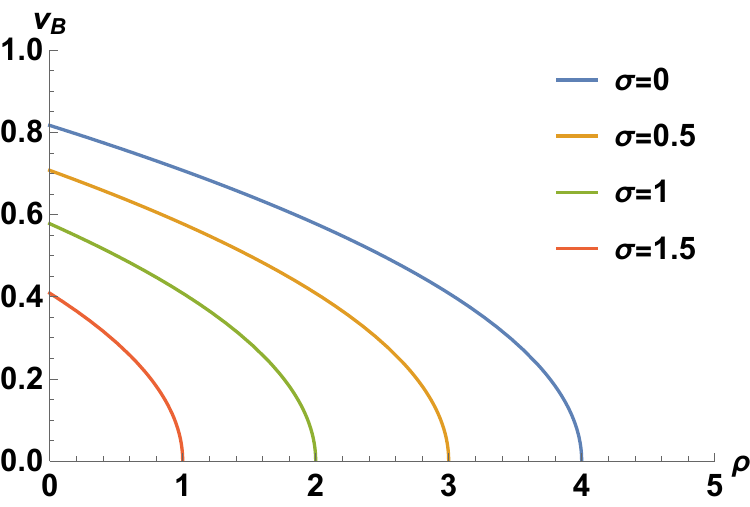}
\caption{Dependence of butterfly velocity $v_B$ on the backreaction parameter $\rho$ at fixed values of the charge parameter $\sigma$. The decrease in $v_B$ with both $\rho$ and $\sigma$ reveals the impact of bulk matter fields and charge on the rate of information scrambling in the dual holographic theory.}
\label{bvplot}
\end{figure}

The exponential form of the OTOC parallels the structure of the shift function \( \alpha(x) \) derived in \cref{alpha-solution} for a shockwave geometry. This correspondence provides a geometric route to extract both the Lyapunov exponent \( \lambda_L \) and the butterfly velocity \( v_B \). Specifically, by matching the form of \( \alpha(x) \propto e^{-M|x|} \) with the spatial dependence of the OTOC in the exponential regime, one obtains:
\[
M = \frac{\lambda_L}{v_B}.
\]
This relation establishes a direct bridge between the spatial profile of bulk gravitational backreaction and the spatiotemporal growth of chaos in the dual field theory. The explicit expressions for the Lyapunov exponent \( \lambda_L \) and the butterfly velocity \( v_B \) can be extracted by matching the exponential growth behavior of the OTOC with the spatial and temporal profile of the shockwave-induced shift function obtained in \cref{alpha-solution} as

\begin{equation}
    \lambda_L = \frac{2\pi}{\beta}, \quad v_B = \frac{2\pi}{\beta M} = \sqrt{\frac{d-\rho-(d-2)\sigma}{2(d-1)}},
\end{equation}
 As a consistency check, we verify that in the simultaneous limit \( \rho \to 0 \) and \( \sigma \to 0 \), the butterfly velocity reduces to its well-known form for the AdS-Schwarzschild black hole
\begin{equation}
    v_B^{\text{AdS}} = \sqrt{\frac{d}{2(d-1)}},
\end{equation}
which is consistent with results obtained in the absence of both backreaction and charge. Furthermore, in the limit \( \sigma \to 0 \), we recover the previously established expression for the butterfly velocity in the holographic string cloud background, as reported in \cite{Chakrabortty:2022kvq}. Below in \cref{bvplot}, we plot $v_B$ as a function of string density $\rho$ for different values of charge parameter $\sigma$.
We can draw the following conclusions from the \cref{bvplot}
 \begin{itemize}
     \item For all values of $\sigma$, the butterfly velocity decreases as the backreaction $\rho$ increases. This indicates that stronger backreaction leads to a slower spread of quantum information in the dual strongly coupled field theory. This suppression of quantum chaos is consistent with the intuitive picture that higher backreaction distorts the bulk geometry more significantly, thereby inhibiting the rate of growth of perturbations. Similar observation documented earlier for string cloud model without charge i.e. $\sigma=0$ in \cite{Chakrabortty:2022kvq}.
     
    \item  Our new finding suggest that at fixed value of $\rho$ the butterfly velocity decreases as $\sigma$ increases. This behavior reflects the fact that charge introduces an effective potential, reducing the efficiency of operator growth in the boundary theory. The presence of charge modifies the near-horizon geometry, leading to a reduction in the scrambling rate characteristic of chaotic dynamics.
     \item  The slope of the decrease in $v_B$ with respect to $\rho$ becomes steeper for higher $\sigma$. This suggests a non-linear coupling between charge and backreaction effects in the bulk, which jointly control the chaotic properties of the theory. The butterfly velocity, being sensitive to the effective geometry near the horizon, captures this intricate interplay, where the combined presence of charge and backreaction significantly affect the information scrambling.
     \item For each fixed value of $\sigma$, the butterfly velocity $v_B$ vanishes at a specific value of the backreaction parameter $\rho$. As $\sigma$ increases, the corresponding value of $\rho$ for which $v_B = 0$ decreases.
  \end{itemize}
  Finally, as a consistency check, we note that the butterfly velocity in the limit of vanishing backreaction and charge approaches $\sim 0.82$, which coincides with the known value of $v_B$ for the five-dimensional AdS-Schwarzschild black hole. In the next section, we turn our attention to the HTMI, which quantifies the total correlation between two subsystems of a thermofield double (TFD) state.



\section{Holographic Thermo Mutual Information}
\label{htmi}
In this section, we investigate the influence of charge on the \emph{Thermo Mutual Information} (TMI) between two subsystems situated in two non-interacting, identical boundary field theories that jointly define a TFD state. The HTMI quantifies correlations between subsystems \( A \) and \( B \) residing on the left and right boundaries of an eternal AdS black hole geometry, which is dual to the TFD state in the holographic context.

The holographic TMI is constructed as a specific linear combination of HEEs, analogous in form to the mutual information \cite{Roberts:2014isa}. For a bipartite system it is defined as
\begin{equation}
    I(A, B) = S(A) + S(B) - S_T(A \cup B),
\end{equation}
where \( S(A) \) and \( S(B) \) are the entanglement entropies of the subsystems \( A \) and \( B \). The third term \( S_T(A \cup B) \) denotes the entanglement entropy of \( A \cup B \) computed in the full TFD state. It is important to note that \( A \) and \( B \) reside in separate, non-interacting copies of the boundary theory. In the holographic picture, $A$ and $B$ are located at the asymptotically left (\( L \))  and right (\( R \) ) boundaries respectively as illustrated in \cref{eternalBH}. Following the standard prescription used for HEE and HMI, we model \( A \) and \( B \) as rectangular strips of equal width \( l \) \cite{Roberts:2014isa,Leichenauer:2014nxa}. The entanglement entropies \( S(A) \) and \( S(B) \) have already been computed in \cref{sec:hee}. Therefore, our focus in this section will be on computing \( S_T(A \cup B) \), which differs from the connected region entanglement entropy \( S(A \cup B) \) considered in the context of HMI in \cref{MI-EWCS}, due to the inter-boundary nature of the TFD state. 

To proceed, we first clarify the extremal surfaces involved in computing $S_T(A \cup B)$. There are two competing candidate surfaces and the entanglement entropy $S_T(A \cup B)$ is determined by the one with minimal area. The first candidate is simply the disconnected configuration: the sum of two individual RT surfaces associated with the regions $A$ and $B$ independently. The second candidate is a connected surface where the boundary of region $A$ is joined with the boundary of region $B$. The disconnected configuration leads to a vanishing TMI, while a non-zero TMI arises from the connected (or ``wormhole'') configuration. This connected surface, commonly referred to as the wormhole surface, spans between the two asymptotic boundaries and passes through the bifurcation surface of the deformed charged eternal black hole. Its embeddings are given by $(t = 0, z, x = -l/2,\ -L/2 \leq x^j \leq L/2)$ and $(t = 0, z, x = l/2,\ -L/2 \leq x^j \leq L/2)$, respectively. This geometric setup is illustrated in \cref{eternalBH} (left). Now to obtain $S(A\cup B)$ we can use the HRT-prescription \cite {Hubeny:2007xt}. With help of induced metrics and the fact that $dx/dz=0$ i.e. for wormhole surfaces we do not have turning point and thus we can write the holographic form of TMI as
\begin{equation}\label{MI-wormhole}
    I_0(A,B)
=
    \frac{
    L^{d-2}R^{d-1}_{AdS}}{G_N^{d+1}} 
    \left[
    \int_{0}^{z_t}\frac{dz}{z^{d-1}\sqrt{f(z)}\sqrt{1-\left(\frac{z}{z_t} \right)^{2d-2}}}-\int_{0}^{z_h}\frac{dz}{z^{d-1}\sqrt{f(z)}}
    \right].
\end{equation}
Here $z_t$ is the turning point of RT surfaces corresponding to the subsystems $A$ and $B$. The turning point $z_t$ can be written in terms of the subsystem width $l$ through the relation \cref{turning-point}.

\begin{figure}[H]
\centering
\includegraphics[width=.40\linewidth]{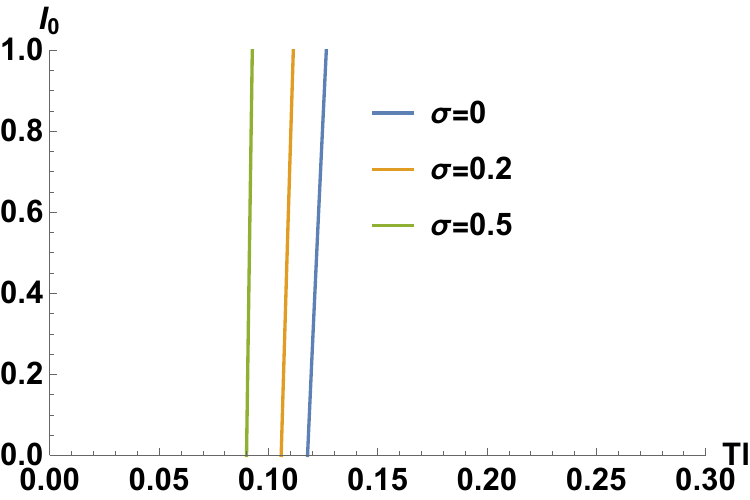}
\caption{TMI as a function of subsystem width \( l \) for fixed temperature \( T = 1 \) and backreaction parameter \( \rho = 1 \), plotted for different values of the charge parameter \( \sigma \). The onset of non-zero TMI occurs beyond a critical width \( l_c \), which increases with \( \sigma \), indicating that charge suppresses inter-boundary correlations.}
\label{HTMI-plot}
\end{figure}

We have plotted the behavior of TMI with respect to the subsystem size $l$ in $d=4$ at a constant temperature $T=1$, which is shown in \cref{HTMI-plot}. Note that there exists a size $l_{c}$ below which the HTMI becomes zero and we will only consider $l\geq l_c$.
In the above figure, we plot the TMI as a function of the subsystem width \( l \) at fixed temperature \( T = 1 \) and backreaction parameter \( \rho = 1 \), for different values of the charge parameter \( \sigma \). The following technical conclusions can be drawn:

\begin{itemize}
    \item  For all values of \( \sigma \), the TMI exhibits a sharp transition from zero to a finite value as \( l \) increases beyond a critical threshold \( l_c \). This marks a transition in the entanglement structure, corresponding to a change in the dominant bulk extremal surface from disconnected to connected configurations.

    \item  As the \( \sigma \) increases, the critical width \( l_c \) shifts to the smaller values. This indicates that a higher charge suppresses the entanglement more rapidly.

    \item  In the dual gravitational picture, larger values of \( \sigma \) means more deformed wormhole geometry of the eternal AdS black hole. This leads to longer extremal surfaces connecting the two boundaries, increasing their area and hence decreasing the holographic TMI. This reflects a geometric suppression of chaos in the bulk due to the effect of charge.
\end{itemize}
In the next section we study the disruption of TMI due to a small perturbation introduced in the asymptotic past on the boundary of eternal black hole.

\section{Disruption of Thermo Mutual Information due to scrambling}
\label{ShockandMI}
In the previous section, we examined how the presence of charge and backreaction influences the entanglement structure between two non-interacting subsystems by analyzing the two-sided mutual information, also referred to as the thermo-mutual information (TMI).
In this section, we investigate how these two-sided correlations are affected in the presence of a perturbation (shock). Specifically, we carry out a holographic analysis of TMI in a perturbed geometry modeled by a shockwave. As discussed in \cref{sec:Shock-analysis}, the effect of the shockwave is to induce a shift in the Kruskal coordinate $U$ by an amount $\alpha$. This shift parameter $\alpha$ characterizes the strength of the shock: a larger value of $\alpha$ corresponds to a more energetic shockwave, leading to a stronger disruption of correlations across the two boundaries.

The HTMI in the presence of a shock wave can be expressed by introducing the \emph{regularized entanglement entropy}, defined as
\begin{equation}
    S^{\text{reg}}(A \cup B; \alpha) = S_T(A \cup B; \alpha) - S_T(A \cup B; \alpha = 0),
\end{equation}
 The regularization ensures that divergences which are not associated with the shock (i.e., unrelated to $\alpha$) are subtracted, isolating the physical effect of the perturbation.
With this definition, the HTMI can be written as
\begin{equation}
\label{MIreg}
    I(A, B; \alpha) = I(A, B; \alpha = 0) - S^{\text{reg}}(A \cup B; \alpha),
\end{equation}
where $I(A, B; \alpha = 0) \equiv I_0$ denotes the HTMI in the unperturbed background, previously computed in \cref{htmi}.

Since $I_0$ is already known, the remaining task is to compute the regularized entanglement entropy $S^{\text{reg}}(A \cup B; \alpha)$. This requires determining the extremal surfaces corresponding to the time-dependent embeddings:
\begin{align*}
    \{t, z(t),\ x = -l/2,\ -L/2 \leq x^j \leq L/2\}, \\
    \{t, z(t),\ x = +l/2,\ -L/2 \leq x^j \leq L/2\}.
\end{align*}
Following the same procedure as for the HEE, the area functional associated with either of these time-dependent embeddings is given by
\begin{equation}
\mathcal{A}=L^{d-2}R^{d-1}_{AdS} \int{{\frac{dz}{z^{d-1}} \frac{1}{\sqrt{ f(z)+\mathcal{C}^2z^{2d-2}}}}}, \qquad t(z)=\pm \int{\frac{dz}{f(z)\sqrt{1+\mathcal{C}^{-2} f(z)z^{2-2d}}}}.
\label{wharea}
\end{equation}
Here $\mathcal{C}=-\left(\frac{R_{AdS}}{z_{0}}\right)^{d-1}\sqrt{-h(z_{0})}$ is the conserved quantity appearing due to explicit time independence of the area functional.
Now using the prescription outlined in \cite{Shenker:2013yza,Shenker:2014cwa,Roberts:2014isa,Jahnke:2018off,Jahnke:2017iwi,Chakrabortty:2022kvq}, the total area integral is naturally decomposed into three segments (shown in \cref{eternalBH} (right)):

\begin{itemize}
    \item \textbf{Region I:} Extends from the right boundary into the bulk up to the outer horizon.
    \item \textbf{Region II:} Continues from the horizon to the turning point located at $z = z_0$.
    \item \textbf{Region III:} Begins at $z = z_0$ and proceeds toward the left boundary, traversing the interior geometry.
\end{itemize}
Each segment contributes to the total area of the extremal surface and the full configuration captures the effect of the in-falling perturbation on the wormhole geometry and the associated entanglement structure. The area integral divided in regions I, II and III, for the wormhole surface can be written as follows

\begin{equation}
\mathcal{A}_w = 
4 L^{d-2}R^{d-1}_{AdS} \left[\int^{z_h}_{0}{{\frac{dz}{z^{d-1}}\biggl( \frac{1}{\sqrt{ f(z)+\mathcal{C}^2z^{2d-2}}}-\frac{1}{\sqrt{ f(z)}}}}\biggr)+2 \int^{z_0}_{z_h}{{\frac{dz}{z^{d-1}} \frac{1}{\sqrt{ f(z)+\mathcal{C}^2z^{2d-2}}}}}\right],
\end{equation}

As discussed in \cref{sec:Shock-analysis}, the strength of the shockwave is characterized by the amount it shifts the event horizon, quantified by the parameter \( \alpha \). However, there exists an alternative bulk parameter, \( z_0 \), which also encodes the shockwave strength and related to \( \alpha \). 
Due to the technical challenges associated with computing \( \alpha \) directly in the context of regularized HEE and TMI, we choose to use \( z_0 \) as the primary parameter to characterize the shockwave strength in our analysis. The precise functional relation between \( \alpha \) and \( z_0 \) will be discussed explicitly in a subsequent section. The regularized HEE is now given by
$S^{reg}_{A\cup B}(z_0) = \mathcal{A}(\gamma_{\text{wormhole}}) / 4 G_N^{d+1}$.

\begin{figure}[H]
\centering
 \includegraphics[width=.40\linewidth]{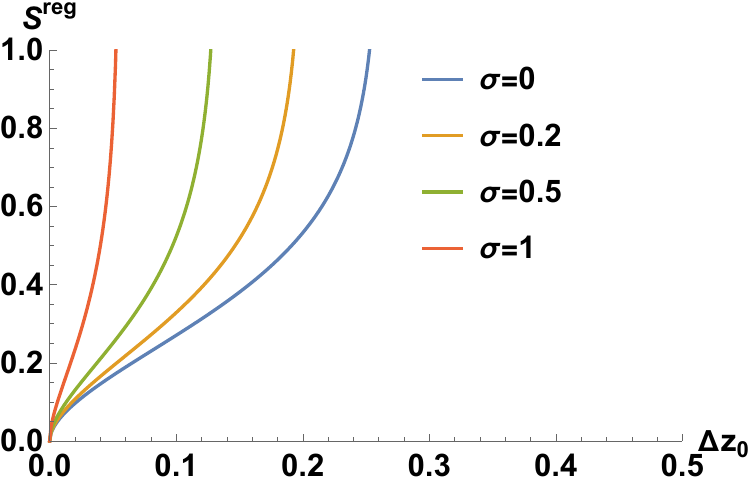}
 \hspace{.10\linewidth}
 \includegraphics[width=.40\linewidth]{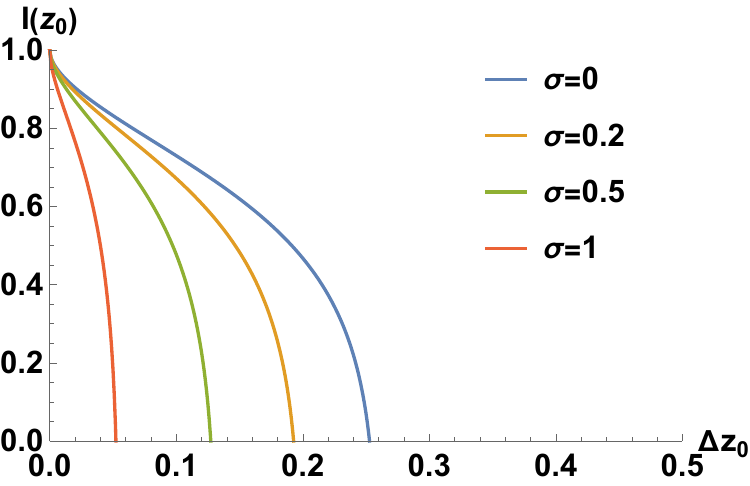}
  \caption{\textbf{Left-} Regularized entanglement entropy \( S^{\text{reg}} \) as a function of the dimensionless shockwave parameter \(\Delta z_0=(z_0-z_h)/z_h \) for fixed backreaction \( \rho = 1,~~z_h=1\) and varying charge parameter \( \sigma \). The increase in \( S^{\text{reg}} \) indicates the increase of inter-boundary entanglement due to shockwave. The increasing rate becomes stronger with increasing \( \sigma \).
  \textbf{Right-} Plot for TMI in presence of shock wave for $d=4, \rho=1$ and different values of $\sigma$. This plot shows that TMI starts from a fixed value and gradually decreases and vanishes for a particular value of $\alpha$.}
\label{Sreg}
\end{figure}

The plot in \cref{Sreg} displays the behavior of the \( S^{\text{reg}}(A \cup B; \alpha) \) and $I(z_0)$ as a function of the dimensionless shockwave parameter \( \Delta z_0 = (z_0 - z_h)/z_h \), for fixed backreaction \( \rho = 1 \), horizon location \( z_h = 1 \), and varying charge parameter \( \sigma \). 
The variable \( z_0 \) which is directly related to the shockwave strength \(\alpha\), parametrizes the turning point of the wormhole surface. A larger value of \( z_0 \) corresponds to a stronger shockwave. The charge \( \sigma \) associated with the chemical potential or electric charge in the dual boundary theory, modulates the background geometry and affects the response of the system to perturbations. Following are the key observations:

\begin{itemize}
    \item For all values of \( \sigma \), \( S^{\text{reg}} \) exhibits increase with increasing \( \Delta z_0 \), signaling progressive increment of inter-boundary entanglement as the shockwave strength increases.
    
    \item The rate of increase is sharper for higher values of \( \sigma \), indicating that the presence of charge enhances the susceptibility of the wormhole geometry to shockwave perturbations. Physically, larger \( \sigma \) corresponds to stronger coupling between the two boundaries due to the additional degrees of freedom introduced by the gauge field.
\item For all values of \( \sigma \), the TMI starts from a non-zero value at small \( \Delta z_0 \), indicating significant inter-boundary correlations in the unperturbed geometry. As \( \Delta z_0 \) increases, the TMI decreases and vanishes at a critical value of \( \Delta z_0 \), beyond which the TMI remains zero. This transition reflects the complete disruption of correlations between the subsystems due to shockwave-induced scrambling.
 \item The critical value of \( \Delta z_0 \) at which TMI vanishes shifts to smaller values as \( \sigma \) increases. This suggests that higher charge enhances the sensitivity of the system to perturbations, i.e., the charged system scrambles information more efficiently.

\end{itemize}
Physically, both the backreaction parameter \( \rho \) and the charge parameter \( \sigma \) introduce additional degrees of freedom into the bulk geometry, modifying the causal structure and facilitating faster delocalization of quantum information. The presence of a shockwave, acting as a localized perturbation, further accelerates this process, consistent with the picture of fast scrambling in holographic systems.

\subsection*{Relation between $\alpha$ and $z_0$:}
We begin by defining the Kruskal coordinates in the three regions of interest, namely Regions~I, II, and III.

\begin{itemize}
    \item \textbf{Region I} spans from the right boundary point at \((\hat{U}, \hat{V}) = (-1, 1)\) to a point on the future horizon at \((\hat{U}, \hat{V}) = (0, \hat{V}_1)\).
    
    \item \textbf{Region II} extends from the horizon point \((\hat{U}, \hat{V}) = (0, \hat{V}_1)\) to the turning point of the extremal surface located at \(z = z_0\), with corresponding Kruskal coordinates \((\hat{U}, \hat{V}) = (\hat{U}_2, \hat{V}_2)\).
    
    \item \textbf{Region III} covers the segment from the turning point \((\hat{U}_2, \hat{V}_2)\) to the left boundary approach, ending at \((\hat{U}, \hat{V}) = (\alpha/2, 0)\).
\end{itemize}
Using the definitions of the Kruskal coordinates as established in Section~\ref{sec:Shock-analysis}, the variation of the Kruskal coordinate \(\hat{V}\) along the extremal surface can be expressed as:
\begin{equation}
\Delta \log \hat{V}^2=\log \hat{V}_1^2-\log{\hat{V}_0^2} =\frac{4\pi}{\beta}(\Delta z_*+\Delta t),~~  \hat{V}_1 =\exp{\biggl[\frac{2\pi}{\beta}\int^{z_h}_{0}{\frac{dz}{f(z)}\biggl({\frac{1}{\sqrt{1+\mathcal{C}^{-2} f(z)z^{2-2d}}}}-1}\biggr)}\biggr].
\end{equation}
In \textbf{Region I}, the radial coordinate increases with the affine parameter, i.e., \( \dot{z} > 0 \), which implies that we should choose the plus (+) sign in the time parametrization. 
In \textbf{Region II}, the blackening function \( f(z) \) becomes negative, which leads to \( \dot{z} < 0 \). Despite this, to maintain the continuity of the embedding and sign conventions in the integral expressions, we again take the plus (+) sign in the time parametrization.
The variation of the Kruskal coordinate \( \hat{V} \) across Region II is then given by
\begin{equation}
\Delta \log \hat{V}^2 = \log \hat{V}_2^2 - \log \hat{V}_1^2 = \frac{4\pi}{\beta}(\Delta z_* + \Delta t),
\end{equation}
where the explicit expression for \( \hat{V}_2 \) reads:
\begin{equation}
\hat{V}_2 = \hat{V}_1 \exp\left[ \frac{2\pi}{\beta} \int_{z_h}^{z_0} \frac{dz}{f(z)} \left( \frac{1}{\sqrt{1 + \mathcal{C}^{-2} f(z) z^{2 - 2d}}} - 1 \right) \right].
\end{equation}
To evaluate \( \hat{U}_2 \), we introduce a reference point at \( \bar{z} \) inside the horizon such that \( z_* = 0 \) at \( z = \bar{z} \). The corresponding Kruskal coordinate is then
\begin{equation}
\hat{U}_2 = \frac{1}{\hat{V}_2} \exp\left[ -\frac{4\pi}{\beta} \int_{\bar{z}}^{z_0} \frac{dz}{f(z)} \right].
\label{u2}
\end{equation}

In \textbf{Region III}, although \( \dot{z} > 0 \), the blackening function \( f(z) \) remains negative. In this case, the appropriate sign choice for time parametrization is negative (-). The variation in the \( \hat{U} \) coordinate across Region III yields
\begin{equation}
\alpha = 2 \hat{U}_2 \exp\left[ \frac{2\pi}{\beta} \int_{z_h}^{z_0} \frac{dz}{f(z)} \left( 1 - \frac{1}{\sqrt{1 + \mathcal{C}^{-2} f(z) z^{2 - 2d}}} \right) \right].
\label{alpha}
\end{equation}
On combining \cref{u2} and \cref{alpha}, we obtain the desired expression relating the shockwave parameter \( \alpha \) to the turning point \( z_0 \) as
\begin{equation}
\alpha(z_0) = 2 \exp\left[ \eta_{\text{I}} + \eta_{\text{II}} + \eta_{\text{III}} \right],
\label{az0}
\end{equation}
where each \( \eta_i \) term encodes the integral contributions from the respective spacetime segments.

\subsection*{Late-Time Growth of Regularized Entanglement and Entanglement Velocity}
Now, we quantify the rate of growth of regularized EE in terms of the \emph{entanglement velocity} \( v_E \), which characterizes the linear growth regime of \( S^{\text{reg}} \) in boundary time \( t_0 \).
To compute the growth rate, we expand \( S^{\text{reg}} \) near \( z_0 \approx z_c \), where \( z_c \) corresponds to the value at which \( \alpha \to \infty \). Expanding to linear order in \( z_0 \), we obtain
\begin{equation}
S^{\text{reg}} \approx \frac{L^{d-2} R^{d-1}_{AdS} \sqrt{-f(z_c)}}{G_N^{d+1} z_c^{d-1}} \cdot \frac{\beta}{4\pi} \log \alpha, \qquad \text{for} \quad z_0 \approx z_c.
\label{srg}
\end{equation}
Since \( \alpha \sim e^{2\pi t_0/\beta} \), \cref{srg} implies that \( S^{\text{reg}} \) grows linearly with boundary time \( t_0 \) in the late-time limit
\begin{equation}\label{a}
\frac{dS^{\text{reg}}}{dt_0} = \frac{2L^{d-2}}{R^{d-1}_{AdS}} \, s \left( \frac{z_h^{d-1}}{z_c^{d-1}} \sqrt{-f(z_c)} \right) = \frac{2L^{d-2}}{R^{d-1}_{AdS}} \, s \, v_E.
\end{equation}
Here, \( s \) denotes the thermal entropy density of the dual field theory and given by 
\begin{equation}
s = \frac{R^{2d-2}_{AdS}}{4G^{d+1}_N} \cdot \frac{1}{z_h^{d-1}},
\end{equation}
From \cref{a}, the entanglement velocity can be identified as
\begin{equation}
v_E = \left( \frac{z_h^{d-1}}{z_c^{d-1}} \right) \sqrt{-f(z_c)}.
\end{equation}

The divergence of \( \alpha \) at \( z_0 = z_c \) is attributed to the divergence of the integral \( \mathcal{\eta}_{\text{III}} \), which captures the contribution from Region III of the extremal surface. Expanding \( \mathcal{\eta}_{\text{III}} \) around \( z = z_0 \) to linear order yields
\begin{equation}
\mathcal{\eta}_{\text{III}} = \frac{4\pi}{\beta} \int_{z_h}^{z_0} \frac{dz}{f(z)} \left[ 1 - \left( -\frac{z_0^{2 - 2d}}{f(z_0)} \left[ f(z) z^{2 - 2d} \right]'_{z=z_0} (z - z_0) \right)^{-1/2} \right].
\end{equation}
The integral \( \mathcal{\eta}_{\text{III}} \) diverges when the coefficient of \( (z - z_0) \) inside the square root vanishes. This occurs when
\begin{equation}
\left. \frac{d}{dz} \left( f(z) z^{2 - 2d} \right) \right|_{z_0 = z_c} = 0.
\end{equation}
This condition determines the critical value \( z_c \), at which the logarithmic growth in \( \alpha \) dominates, setting the late-time linear growth regime of the entanglement entropy.

A detailed analysis reveals that the entanglement velocity \( v_E \) exhibits a qualitative trend analogous to that of the butterfly velocity \( v_B \), as discussed in \cref{sec:Shock-analysis}. For fixed values of the backreaction  \( \rho \) and the charge parameter \( \sigma \), the entanglement velocity remains consistently bounded from above by the butterfly velocity, i.e., \( v_E < v_B \). This inequality, \( v_B \geq v_E \), is a characteristic feature of holographic systems and has been previously reported in \cite{Mezei:2016zxg, Mezei:2016wfz, Chakrabortty:2022kvq}.
Furthermore, we observe that in the presence of non-zero backreaction and charge, the entanglement velocity asymptotically approaches the upper bound \( \sqrt{d/2(d-1)} \), mirroring the saturation behavior noted for \( v_B \). These results are consistent with a broader body of recent work—both numerical and analytic—on the linear growth regime of entanglement entropy in quantum many-body systems following global quenches, particularly within the holographic framework \cite{Liu:2013qca}.

\section{Summary and Discussion}\label{summary}
In this work, we present a detailed numerical study of how charge and backreaction affect various measures of entanglement and chaos. Specifically, we analyzed the HEE, HMI, EWCS, butterfly velocity and holographic TMI in the background of a backreacted charged AdS black hole. This geometry is dual to a strongly coupled large-$\mathcal{N}_c$ field theory at finite temperature and chemical potential, in presence of uniform distribution of heavy static fundamental quarks. The quantities we compute serve as sensitive probes of entanglement structure, correlation strength and information scrambling in the dual theory, offering deep insights into how the bulk geometry encodes and influences boundary dynamics.

Our analysis shows that HEE increases monotonically with subsystem size \( l \), consistent with the expectation that larger subsystems encode more degrees of freedom. The presence of both backreaction (\( \rho \)) and charge (\( \sigma \)) modifies the background geometry by introducing additional bulk energy-momentum content, effectively increasing the entangling surface area and hence the EE. We find that increasing either \( \rho \) or \( \sigma \) independently leads to higher HEE, a result that is consistent with previous studies in the zero-charge limit \cite{Chakrabortty:2020ptb}, and extended here to the charged regime. For holographic MI, we observe that increasing $\rho$ enhances inter-subsystem correlations, suggesting that the additional matter content strengthens long-range entanglement. However, the rate of this enhancement becomes slower  for increasing charge, indicating a competition between the correlation-enhancing effect of backreaction and charge.  
Interestingly, MI vanishes beyond a critical value of \( \rho \), a bound which itself decreases as \( \sigma \) increases, reflecting a charge-induced limitation on correlation strength. The EWCS displays a monotonic increase with both \( \rho \) and \( \sigma \). This trend geometrically reflects the stretching of the entanglement wedge and the enlargement of the minimal cross-sectional area as the bulk geometry becomes more curved and richer in matter content.

In the study of butterfly velocity, we find that increasing backreaction and charge decreases the $v_B$  as a result slows the spread of perturbations. At some critical $\rho$, $v_B$ becomes zero and for large charge less backreaction is required to make the $v_B$ zero, suggesting that charged systems scramble information more rapidly and become more chaotic.  For consistency in the zero charge limit our result is in agreement with previous findings in uncharged deformed AdS black hole \cite{Chakrabortty:2022kvq}. A novel result of our analysis is that, for fixed backreaction, increasing charge suppresses \( v_B \) more rapidly, implying that the gauge sector introduces an effective potential barrier that inhibits the growth of operator size. Moreover, the rate at which \( v_B \) decreases with respect to \( \rho \) becomes steeper as \( \sigma \) increases. 
 
On the other hand TMI, that captures correlations between subsystems in a TFD state, exhibits a sharp transition from zero to non-zero values as the width \( l \) of the entangling region crosses a critical threshold \( l_c \). This behavior reflects a topological transition in the dominant extremal surface configuration from disconnected to connected, consistent with prior holographic studies \cite{Jahnke:2017iwi, Jain:2023xta, Chakrabortty:2022kvq, Karan:2023hfk}. Increasing \( \sigma \) reduces the critical width \( l_c \), implying that charge weakens inter-boundary entanglement and facilitates earlier decoherence.

Finally, we studied the response of holographic TMI under shockwave perturbations by examining the regularized entropy \( S^{\text{reg}} \) as a function of the shockwave strength parameter \( \Delta z_0 \). Our results show that \( S^{\text{reg}} \) increases with \( \Delta z_0 \), reflecting growing entanglement due to the wormhole geometry being increasingly deformed. However, TMI exhibits a monotonic decrease with \( \Delta z_0 \) and vanishes beyond a critical threshold, marking the complete scrambling of correlations. As \( \sigma \) increases, the critical value of \( \Delta z_0 \) at which TMI vanishes decreases, indicating that charge enhances the system’s ability to shockwave-induced disruption and accelerates information scrambling. This behavior highlights the dual role of charge as both an entanglement enhancer and a scrambling catalyst.

Taken together, our results reveal a coherent and physically consistent picture in which backreaction and charge jointly influence the dynamics of entanglement and chaos in holographic theory. While backreaction generally amplifies both correlation and complexity by increasing the matter content in the bulk, charge introduces subtler effects: it enriches the entanglement structure while simultaneously screening correlations and suppressing operator growth. The nonlinear interplay between these parameters controls the response of dual field theory to the perturbation, and strongly affects its chaotic and entanglement structure.

Several extensions of this work merit further exploration. A natural next step is the analysis of time-dependent perturbations and entanglement growth following global or local quenches in the presence of charge and backreaction. It would also be interesting to investigate the analytic behavior of entanglement measures in this background similar to \cite{Chakrabortty:2020ptb}, that may offer new insights into the structure of mixed-state correlations. Finally, from a boundary field theory perspective, understanding how chemical potential, or finite density affects scrambling and mutual information remains an important open problem.

\section*{Acknowledgments}

A.M. would like to thank the council of Scientific and Industrial Research (CSIR), Government of India, for the financial support through a research fellowship (File No.:09/1005(0034)/2020-EMR-I). S.P. gratefully acknowledges Shankhadeep Chakrabortty for his valuable support during the visit to IIT Ropar. S.P. also thanks Graphic Era for their institutional support in facilitating this work. H.P. acknowledges the support of this work by NCTS.

\bibliographystyle{JHEP}
\bibliography{ChaosC}

\end{document}